\documentclass[structabstract]{aa}  
\usepackage{amsmath}

\usepackage{graphicx}
\usepackage{txfonts}
\usepackage{natbib}

\usepackage{color}

\usepackage{comment}

\usepackage{soul}

\newcommand {\nbody} {\textsc{\mbox{nbody6}}}
\newcommand {\nbodypp} {\textsc{\mbox{nbody6\raise.4ex\hbox{\tiny++}}}}
\newcommand {\secnbodypp} {\textsc{\mbox{nbody6\raise.4ex\hbox{\scriptsize++}}}}

\newcommand {\COri} {\mbox{$\theta^1\mathrm{C}\;\mathrm{Ori}$}}
\newcommand {\Msun} {\mbox{M$_{\odot}$}}

\newcommand {\days} {\mbox{d}}
\newcommand {\Myr} {\mbox{Myr}}

\newcommand {\AU} {\mbox{AU}}
\newcommand {\pc} {\mbox{pc}}

\newcommand {\kms} {\mbox{km\,s$^{-1}$}}
\newcommand {\pcdens} {\mbox{pc$^{-3}$}}

\newcommand {\Rhm} {\mbox{R$_{\textrm{hm}}$}}
\newcommand {\binaryfrequency} {\mbox{$b_f$}}

\newcommand {\Nbin} {\mbox{$N_b$}}
\newcommand {\Nbinnorm} {\mbox{$\mathcal{N}_b$}}

\newcommand {\Mprim} {\mbox{$M_{\mathrm{prim}}$}}

\newcommand {\Msec} {\mbox{$M_{\mathrm{sec}}$}}

\newcommand {\Nsing} {\mbox{$N_s$}}
\newcommand {\Nsingnorm} {\mbox{$\mathcal{N}_s$}}

\begin{document}
   \title{Evolution of the binary population in young dense star clusters}
   %\titlerunning{Self-similar evolution of binary populations in star clusters}

   %\subtitle{}

   \author{T. Kaczmarek\inst{1} \and C. Olczak\inst{2,}\inst{3,}\inst{4,}\inst{5} \and S. Pfalzner\inst{6}
          }

   \institute{
     \inst{1}I. Physikalisches Institut, Universit\"at zu K\"oln, Z\"ulpicher Str. 77, 50937 K\"oln, Germany\\
     \inst{2}Astronomisches Rechen-Institut (ARI), Zentrum f\"ur Astronomie der Universitaet Heidelberg (ZAH) M\"onchhofstr. 12-14,
     69120 Heidelberg, Germany \\
     \inst{3}Max-Planck-Institut f\"ur Astronomie (MPIA), K\"onigstuhl 17, 69117 Heidelberg, Germany\\
     \inst{4}National Astronomical Observatories of China (NAOC), Chinese Academy of Sciences (CAS) 20A Datun Lu, Chaoyang
     District, Beijing 100012, China\\
     \inst{5}The Kavli Institute for Astronomy and Astrophysics (KIAA), Peking University (PKU) Yi He Yuan Lu 5, Hai Dian Qu,
     Beijing 100871, China\\
     \inst{6}Max-Planck-Institut f\"ur Radioastronomie (MPIfR), Auf dem H\"ugel 69, 53121 Bonn, Germany\\
     \email{kaczmarek@ph1.uni-koeln.de}
   }

   \date{}

% \abstract{}{}{}{}{} 
% 5 {} token are mandatory

   \abstract
% context heading (optional)
% {} leave it empty if necessary  
{ Field stars are not always single stars, but can often be found in bound double systems. Since binary frequencies in the birth
  places of stars, young embedded clusters, are sometimes even higher than on average the question arises of how binary stars form
  in young dense star clusters and how their properties evolve to those observed in the field population.  }
% aims heading (mandatory)
{We assess, the influence of stellar dynamical interactions on the primordial binary population in young dense cluster environments.}
% methods heading (mandatory)
{ We perform numerical N-body simulations of the Orion Nebula Cluster like star cluster models including primordial binary
  populations using the simulation code \nbodypp\.}
% results heading (mandatory)
{ We find two remarkable results that have yet not been reported: The first is that the evolution of the binary frequency in young
  dense star clusters is independent predictably of its initial value. The time evolution of the normalized number of binary
  systems has a fundamental shape. The second main result is that the mass of the primary star is of vital importance to the
  evolution of the binary.  The more massive a primary star, the lower the probability that the binary is destroyed by
  gravitational interactions. This results in a higher binary frequency for stars more massive than $2\Msun$ compared to the
  binary frequency of lower mass stars.  The observed increase in the binary frequency with primary mass is therefore most likely
  not due to differences in the formation process but can be entirely explained as a dynamical effect.  }
% conclusions heading (optional), leave it empty if necessary 
{ Our results allow us to draw conclusions about the past and the future number of binary systems in young dense star clusters and
  demonstrate that the present field stellar population has been influenced significantly by its natal environments.}


   \maketitle
%
%________________________________________________________________

\section{Introduction}

Today the formation process of isolated single stars seems to be understood in principle. One remaining uncertainty, at least to
some extent, concerns the formation of binary stars. There are several ways to form binary stars - the fragmentation of a collapsing
cloud or core or the dynamical capture of a companion.  However, determining the dominating process of binary formation requires
knowledge of the parameters of the primordial binary population, i.e. the population as it appears just after star formation has
finished.

Several surveys have been performed to determine the present-day parameters of binary populations in different kinds of
environments. Observations of the field binary population have shown that the binary frequency is quite high, $60\%$ for G-type
dwarf stars \citep{1991A&A...248..485D} and $35 - 42\%$ for M-type dwarfs \citep{1992ApJ...396..178F,1997AJ....113.2246R}. Both
\cite{1991A&A...248..485D} and \cite{1992ApJ...396..178F} determined the mass ratio and separation/period distributions of their
samples. They found that the mass ratio distribution for G-type dwarfs increases as the  mass ratios decreases while the mass ratio
distribution for M-dwarfs is flat. The separation/period distributions for G-type and M-type dwarfs can be fitted by a Gaussian
distribution in logarithmic space where both peak at nearly the same separation ($\sim 30 \AU$ for G-type and $9-30\AU$ for M-type
dwarfs) with the FWHM of both distributions spanning over three orders of magnitude \citep[see][Fig.~7]{1991A&A...248..485D}.

Another feature of binary populations is their dependence on the primary mass: observations indicate that the probability of finding
a massive star in a binary system is higher than for low-mass stars.  For OB stars, a binary frequency of $\sim 70\%$ is found
\citep{2002AJ....124..404P}, whereas for G-dwarfs a binary frequency of $\sim 60\%$ \citep{1991A&A...248..485D} and for M-dwarfs
a value as low as $\sim 31\%$ \citep{2003IAUS..211..311M} has been observed.

An understanding of the presently observed properties of evolved binary populations requires knowledge of the primordial binary
population.  To achieve this one has to look at sites where star formation has just happened, such as young star
clusters. However, observations of binary populations in star clusters do not provide such a clear picture as the observations of the
field binary population. While in dense star clusters such as the Orion Nebula Cluster (ONC) or NGC 2024 the observed binary
frequency is similar to the field binary frequency \citep{2006A&A...458..461K,2003ApJ...583..358B}, sparse clusters such as the
Taurus-Auriga association have much higher binary frequencies \citep[see][and references therein]{1999A&A...341..547D}. This leads
to the question of what defines the binary frequency in the different environments. Possible explanations are differences in the
temperatures of the parent clouds \citep{1994A&A...286...84D} or the dynamical evolution within the cluster
\citep[e.g.][]{1999NewA....4..495K}. The latter possibility will be investigated in this study.

According to current knowledge, stars do not form in isolation, but in a clustered environment consisting of several tens to
thousands of stars. Most of them possibly formed in clusters such as the ONC or even denser environments
\citep{2003ARA&A..41...57L,2009A&A...498L..37P}. The stellar density in ONC-like clusters is sufficiently high to potentially
affect the evolution of the binary population via gravitational interactions.

The gravitational interactions that alter binaries in a star cluster are mostly three- and four-body interactions between binary
and single stars. \citet{1975MNRAS.173..729H,1975IAUS...69...73H} and \citet{1975AJ.....80..809H} showed that three-body
interactions are very effective in altering the parameters of binary systems.  Both authors found that loosely bound binaries
(also called soft) tend to become softer (meaning that their internal energy increases resulting in the widening of the binaries)
owing to three-body interactions, while firmly bound (hard) binaries tend to become harder (Heggie-Hills law).

Four-body interactions (binary-binary encounters) become important in clusters that consist of
a high number of binary stars \citep[see for example Fig.~1 in ][]{2010MNRAS.tmp.1645L}. In addition to three-body interactions,
four-body interactions can form two ``new'' binary systems by exchanging the binary components, can destroy one or even both
binary systems, or can lead to the formation of bound hierarchies of three or more stars. \nbodypp
\citep{2003gnbs.book.....A,1999JCoAM.109..407S}, the simulation code used for this study, uses several techniques such as the
regularization of three-body interactions \citep{1974CeMec..10..185A} or the chain method \citep{1993CeMDA..57..439M} to compute
the evolution of these interactions correctly. However, in this study we only diagnose the evolution of the binary population.

Numerical investigations of the dynamical evolution of binaries in star clusters started in the 70's with the development
of the first N-body codes \citep[e.g.][]{1971Ap&SS..13..324A}.  The early works concentrated on the formation of binaries in star
clusters with initially pure single stellar systems.  Only in recent years simulations have been performed to investigate the
evolution of a primordial binary population. For example, \citet{1999NewA....4..495K} simulated the evolution of ONC-like star
cluster models where all stars initially were part of a binary system, i.e.  a primordial binary frequency of $100\%$. They showed
that the cluster evolution destroys wide binaries quite efficiently leading to a reduction in the initial binary fraction as the
cluster evolves.

Here we also want to study the evolution of the primordial binary population of a young dense star cluster. In contrast to the
previous studies, we investigate the effect of the initial binary frequency ($0-100\%$) and a log-uniform semi-major axis
distribution similar to that observed in very young star forming regions
\citep[e.g.][]{2007AJ....134.2272R,2008AJ....135.2526C}. Because observations indicate that the binary frequency increases with
stellar mass, we also focus on the dependence of the binary evolution on stellar mass.
 
This paper is structured in the following way. In Sec.~\ref{sec:definitions} we define parameters that describe a binary
population. Sec.~\ref{sec:method} covers the setup of the simulations, including the generation of the cluster models and
initialization of the primordial binary population. Finally in Sec.~\ref{sec:results} we present the results obtained from the
simulations.

%__________________________________________________________________

\section{Definitions}
\label{sec:definitions}

The most simple measure of the binary population of a star cluster is the number of binaries (\Nbin) within the cluster. Here, a
pair of stars is called a binary system if the system is bound and the two stars are at the same time nearest neighbors.  In
this definition, a star that is part of a multiple system can be part of several ``binary'' systems simultaneously since two stars
do not have to be mutually nearest neighbors. Hence triple or other higher order multiples are not handled separately by our
diagnostic but are counted as multiple binary systems.  In this study, we mostly use the normalized number of binary systems
\Nbinnorm\ defined as
\begin{equation}
  \label{eq:nbin_norm}
  \Nbinnorm (t) = \frac{\Nbin (t)}{\Nbin (0)},
\end{equation}
where $\Nbin (0)$ denotes the initial and $\Nbin (t)$ the number of binary systems at time $t$ within the cluster. The usual
measure of binary populations in star clusters is the binary frequency \binaryfrequency\ 
\begin{equation}
  \label{eq:binary_frequency}
  \binaryfrequency (t) = \frac{\Nbin (t)}{\Nsing (t)+\Nbin (t)},
\end{equation}
where $\Nsing (t)$ denotes the actual number of observed single stars and $\Nbin (t)$ the actual number of binary systems. One
major result of our work is that the normalized number of binary systems, Eq.~\ref{eq:nbin_norm}, is the appropriate measure of a
binary population.

%
%______________________________________________________________

\section{Method and cluster setup}
\label{sec:method}

Here the evolution of binary populations in young dense star clusters is studied using the parallized version \nbodypp\ of the
high-accuracy N-body code \nbody\ \citep{2003gnbs.book.....A,1999JCoAM.109..407S}. The study presented can be understood as a
follow up of \citet{2007A&A...475..875P}.  The main difference to \citet{2007A&A...475..875P} is that here the clusters contain a
primordial binary population, with varying initial binary frequencies. In both studies, the ONC has been chosen as a model cluster
because it is one of the most well-studied star forming regions for which many parameters are well constrained \citep[see for
example][]{1994ApJ...421..517P,1997AJ....113.1733H,1998ApJ...492..540H}, in particular several surveys of its binary population
\citep{1998ApJ...500..825P,1999NewA....4..531P,2006A&A...458..461K,2007AJ....134.2272R} exist.  This allows a detailed comparison
of the simulated binary populations with the observed ones.

Our simulations started with 4000 stellar systems, where stellar systems correspond to either single stars or binary
systems. This means that for a simulation with a primordial binary frequency of 100\% the dynamical evolution of 8000 stars is
calculated. The masses $M$ of the 4000 stellar systems were sampled from the initial mass function given by
\citet{2002Sci...295...82K}
\begin{equation}
  \label{eq:kroupa-imf}
  \xi(M) =
  \begin{cases}
    \; M^{-1.3} \quad & \textrm{if} \quad 0.08 \leq M \left[ \Msun \right]< 0.5, \\
    \; M^{-2.3} \quad & \textrm{if} \quad 0.5 \leq M  \left[ \Msun \right]< 1.0, \\
    \; M^{-2.3} \quad & \textrm{if} \quad 1.0 \leq M  \left[ \Msun \right]< \infty.
  \end{cases}
\end{equation}
The minimum and maximum masses of the stellar systems have been chosen as $M_{\textrm{min}}=0.08 \Msun$, corresponding to the
hydrogen-burning limit, and $M_{\textrm{max}}=50\Msun$ the mass of the most massive stellar system in the ONC - \COri.

Observations of the stellar content in the ONC have shown that the stellar radial density profile of the cluster can be
approximated by an isothermal sphere (i.e. $\rho_{\mathrm{present}} \propto 1/r^{2}$, see
e.g. \cite{1988AJ.....95.1755J,1997AJ....113.1733H}) with the exception of the cluster core ($R_{\mathrm{core}}\approx 0.2\pc$),
which has a much flatter density profile ($\rho_{\mathrm{present,core}} \propto r^{-0.5}$, \citet{2005MNRAS.358..742S}). To
account for this, we adopt an initial density profile of the form
\begin{equation}
  \label{eq:initial_radial_density_profile}
  \rho_{0} (r) = 
  \begin{cases}
    \rho_{0}\; (r/R_{\mathrm{core}} )^{-2.3}&,\quad r\in (0,R_{\mathrm{core}}] \\
    \rho_{0}\; (r/R_{\mathrm{core}} )^{-2.0}&,\quad r\in (R_{\mathrm{core}}, R]\\
    \qquad 0&, \quad r\in (R,\infty]
  \end{cases}
\end{equation}
where $\rho_{0} = 3.1\times 10^{3}\pcdens$, $R_{\mathrm{core}} = 0.2\pc$, and $R=2.9\pc$. \citet{2010A&A...509A..63O} demonstrated
that this initial radial density profile evolves towards the currently observed one after about $1\Myr$ of dynamical evolution,
the estimated age of the ONC \citep{1997AJ....113.1733H}.

Observations of the ONC indicate that the cluster are mass-segregated \citep[e.g.][]{1997AJ....113.1733H}. Numerical simulations
by \citet{1998MNRAS.295..691B} comparable to the cluster models studied here demonstrated that this mass segregation cannot be
due to dynamical evolution of the cluster and therefore must be primordial. Following their suggestion, we initially place the most
massive stellar system at the cluster centre and the three next most massive stars at random positions within a sphere of radius
$R=0.6\Rhm$ centred on the cluster centre, where \Rhm\ is the half-mass radius of the cluster.

The velocities of the stellar systems are sampled from a Maxwellian velocity distribution in virial equilibrium.  The resulting
velocity dispersion of our model is lower than the observational value \citep[$\sigma = 4.3\pm
0.5\kms$:][]{1988AJ.....95.1755J}. However, because of significant mass of the molecular cloud associated with the ONC that was
neglected in our numerical model, the observed velocity dispersion is consistent without any assumption about a stellar system in
dynamical equilibrium. In future investigations of the binary populations, we plan to include the effect of the gas associated
with the cluster.

For the primordial binary population, we chose the model described in \citet{2007A&A...474...77K}.  They compared observations of
the Scorpius OB II association with simulated observations of their binary population model to eliminate observational selection
effects and biases that affect real observations of star clusters. Their analysis shows that the binary frequency in Scorpius OB
II is above $70\%$ with a confidence of $3\sigma$. The closest agreement with observations is obtained for models with a binary
frequency of $100\%$. The model of \citet{2007A&A...474...77K} was chosen here since it is a well-studied binary population, close
to its initial conditions. Scorpius OB II is a relatively young (about $5-20\Myr$) and probably too sparse region to have
undergone significant dynamical evolution.

In contrast to the field population that exhibits a log-normal period distribution of G and M-dwarf binaries
\citep{1991A&A...248..485D,1997AJ....113.2246R}, the younger population in Sco OB II has a flat distribution of the semi-major
axis in logarithmic space \citep{2007A&A...474...77K}
\begin{equation}
  \label{eq:kouwenhoven07-oepiks-law}
  f_a(a) \propto a ^{-1} \qquad \Leftrightarrow \qquad f_{\log_{10} a} (\log_{10} a) = \mathrm{const.},
\end{equation}
where $a$ is the semi-major axis of the binary system. 

A flat distribution of the semi-major axes in logarithmic space corresponds to a similar distribution of periods $P$ with the
relation
\begin{align}
  \label{eq:kouwenhoven07-period-separation-distribution}
  \overline{ \log_{10} a } & =  \frac{2}{3}\overline{ \log_{10} P } - \frac{1}{3} \log\left( \frac{4\pi^2}{G\;M_T}\right), \\
  \sigma_{\log_{10} a} & = \frac{2}{3}\sigma_{\log_{10} P},
\end{align}
where $G$ is the gravitational constant and $M_T$ the total mass of a binary system.  

A flat binary period distribution has not only been observed in Sco OB II but also in other regions
\citep[e.g.][]{2008AJ....135.2526C,2007AJ....134.2272R} and has been used in several theoretical studies involving the evolution
of primordial binary frequencies \citep{1992MNRAS.257..513H,2005MNRAS.358..572I,2007ApJ...665..707H}.

In addition to a log-uniform shape of the initial period distribution, other distributions have been used in numerical studies of
the dynamical evolution of binary populations. For example, \citet{1995MNRAS.277.1491K} demonstrated that a dynamically evolved
period distribution that is initially rising for short periods and nearly constant for long periods matches the observed period
distribution of the field G-type dwarfs \citep{1991A&A...248..485D}. In contrast, this cannot be achieved with an initially
log-uniform period distribution because of the much larger number of dynamically stable short-period binaries. However, physical
processes that were not taken into account by \citet{1995MNRAS.277.1491K} could significantly lower the number of short period
binaries. For example ``the orbital decay of embedded binary stars'' \citep{2010MNRAS.402.1758S} is capable of reshaping a
log-uniform binary period distribution within less than 1 \Myr\ towards a log-normal distribution (H\"ovel et al., in
preparation). Consequently, as the effect of different initial period distributions on the evolution of the binary population in
star clusters is basically unknown we will explicitly focus on this topic in a future investigation. In this context the present
study can be understood as preparatory work.

\begin{figure}
  \centering
  \includegraphics[width=\linewidth]{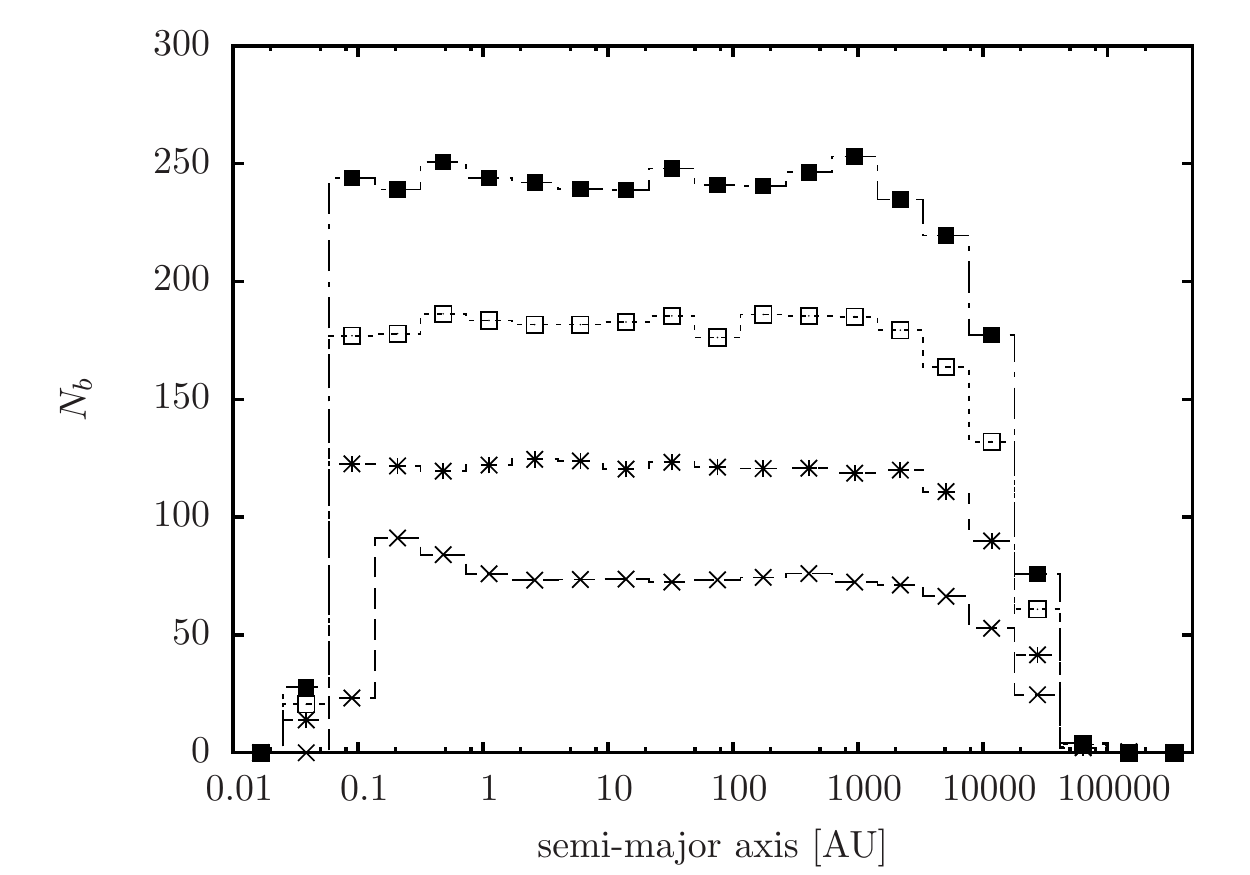}
  \caption{Initial semi-major axis distribution for all simulated clusters. Crosses, stars, open and filled squares denote
    simulations with initially 30\%, 50\%, 75\%, and 100\% binaries, respectively.}
  \label{fig:initial_semi-major_axis_distribution}
\end{figure}

Figure~\ref{fig:initial_semi-major_axis_distribution} shows the initial semi-major axis distribution of our simulations.  Although
the semi-major axis distributions have been set up log-uniformly in the range $4.6\times 10^{-2} \AU\; - \; 4.6\times 10^4 \AU$,
the distribution is only flat up to about $2\times 10^3\AU$ and declines afterwards to zero. The apparent lack of wide binary
systems in the $2\times 10^3 - 4.6\times 10^4\AU$ range is due to the generating algorithm. In the code, binary systems are formed
in isolation and then placed in the cluster environment. The wider a binary, the higher the probability that (at least) one of its
components has a closer neighbour among the other cluster members. According to our definitions (see Sec.~\ref{sec:definitions}),
this chance neighbour is now the potential companion and most probably not bound.

\begin{figure}
  \centering
  \includegraphics[width=\linewidth]{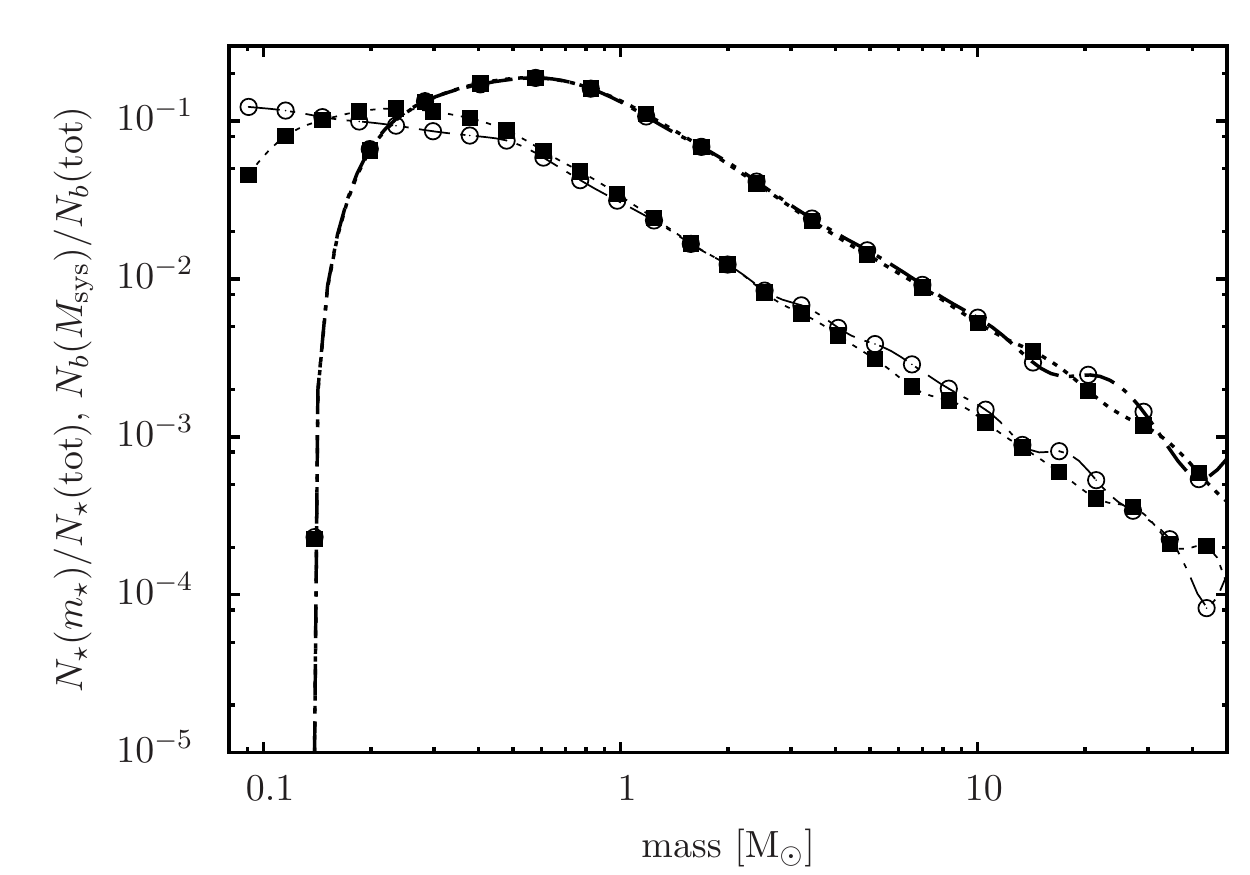}
  \caption{Comparison of the stellar initial mass function (thin lines) and the initial binary-system mass function (thick
    lines). The open circles denote simulations that used random pairing to pair the binary components while the solid squares
    denotes simulations that used Eq.~\ref{eq:kouwenhoven07-mass-ratio-distribution}.}
  \label{fig:comparison_imf_kroupa}
\end{figure}

\citet{2007A&A...474...77K} also found the mass ratio distribution to have the form
\begin{equation}
  \label{eq:kouwenhoven07-mass-ratio-distribution}
  f_q(q)\propto q^{\gamma_q}, \qquad \gamma_q \sim -0.4,
\end{equation}
where $q = \Msec / \Mprim$ is the mass-ratio of a binary with primary mass $\Mprim$ and secondary mass $\Msec$. Adopting this
functional form in our models, the binary system masses were sampled from the \citet{2002Sci...295...82K} IMF and then divided
using Eq.~\ref{eq:kouwenhoven07-mass-ratio-distribution}. This approach, however, does not reproduce the observed stellar IMF as
can be seen in Fig.~\ref{fig:comparison_imf_kroupa} (thin lines), which displays the stellar IMFs produced in our simulations
(filled squares, representing simulation with initially $100\%$ binaries). For comparison, the standard IMF was included by
setting up simulations using random pairing of the binary components instead of using
Eq.~\ref{eq:kouwenhoven07-mass-ratio-distribution} to pair the binary components. Random pairing has been shown to reproduce the
stellar IMF \citep{1993A&A...278..129L} and is here therefore used to represent the standard IMF. Using
Eq.~\ref{eq:kouwenhoven07-mass-ratio-distribution} to pair the binary components generates too few low-mass stars in the range
$0.08 \Msun - 0.16\Msun$ and too many intermediate mass stars in the mass range $0.15 - 0.5\Msun$, where the deviations become
larger the higher the initial binary frequency is. The number of stars generated with masses exceeding $0.5\Msun$ reproduces the
\citet{2002Sci...295...82K} IMF quite well, although their numbers are slightly under-predicted.

The thick lines in Fig.~\ref{fig:comparison_imf_kroupa} show the initial binary-system mass ($m_{\mathrm{Hy's} = \Mprim + \Msec}$)
functions for the simulations used in this study and additionally the initial binary-system mass function if the binary components
are drawn randomly from the stellar IMF. The binary system mass function in both cases is very similar. We do not expect the
differences in the single-star IMFs to affect the evolution of the binary populations because it depends on the system masses as
we show in Sec.~\ref{sec:results}.

The eccentricity of the binary systems in Sco OBII was found by \citet{2007A&A...474...77K} to be thermally distributed,
in agreement with the eccentricities of long period binaries in the field. Following \citet{2001ApJ...555..945K} who illustrated
that the dynamical evolution is not capable of changing the eccentricity distribution significantly, a thermal period distribution
was applied to all of our simulations.

Observations show that the binary frequency varies among low-mass stars in the field ($35-42\%$)
\citep{1992ApJ...396..178F,1997AJ....113.2246R}, Sun-like stars in the solar neighbourhood ($\sim 60\%$)
\citep{1991A&A...248..485D}, and the high binary frequency in low-mass star-forming regions such as Taurus ($\sim 100\%$)
\citep{1999A&A...341..547D}. We account for the large spread among the different environments by investigating cluster models
initially with $30\%$, $50\%$, $75\%$, and $100\%$ binary frequency.
 
To improve the statistical significance of our results, for each initial binary frequency 100 simulations were
performed. Only those that fulfilled the following conditions were considered in the subsequent diagnostic process:
\begin{itemize}
\item after 1 \Myr\ of cluster evolution, the most massive stellar system in the simulations has to be within $1\pc$ of the
  cluster centre.
\item the cluster density profile after 1 \Myr\ of cluster evolution has to agree with the cluster density profiles observed by
  \cite{1997AJ....113.1733H} and \citet{2002Msngr.109...28M}.
\end{itemize}

%
%______________________________________________________________

\section{Results}
\label{sec:results}

\begin{figure}
  \centering
  \includegraphics[width=\linewidth]{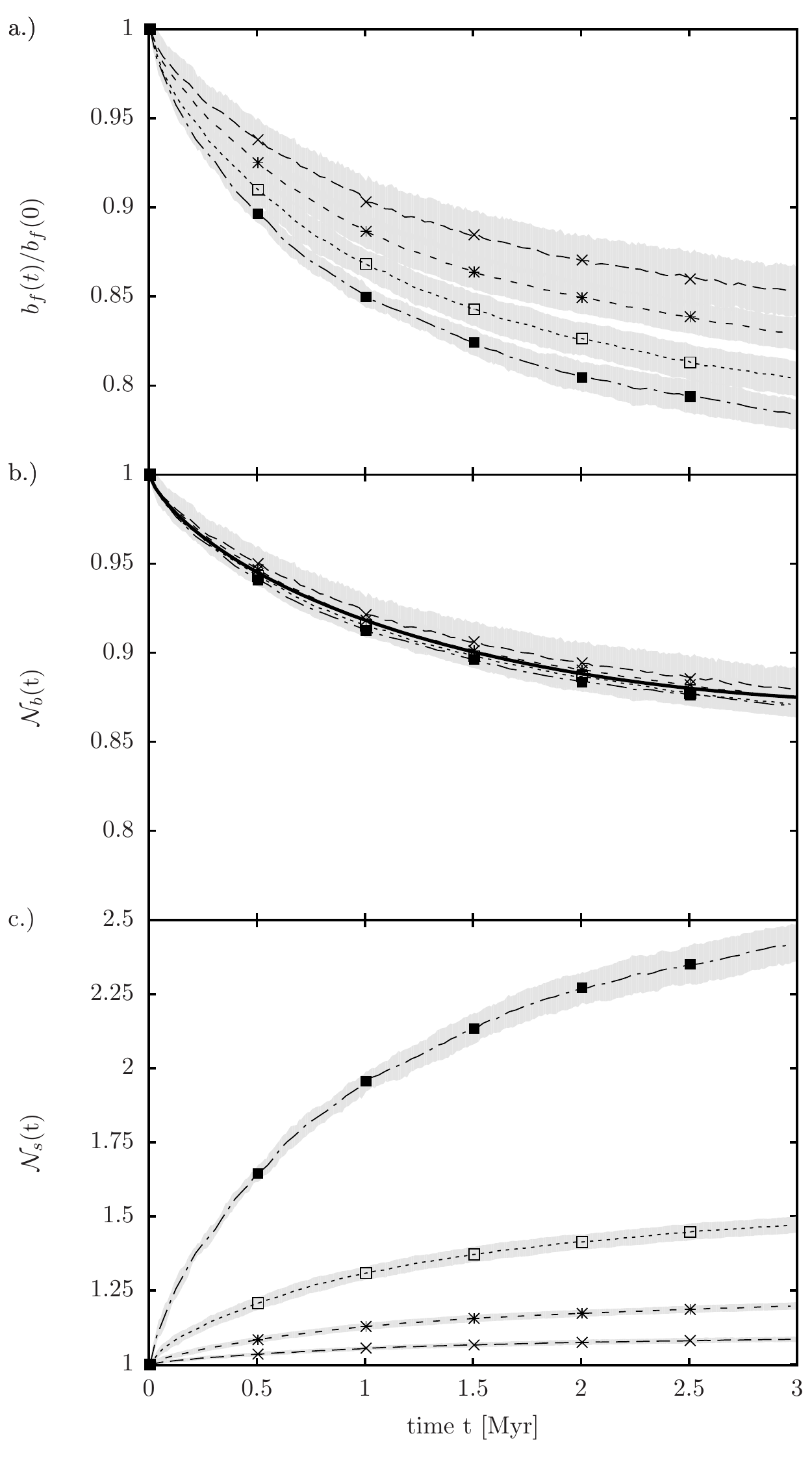}
  \caption{Evolution of the normalized binary frequency, normalized number of binary systems and normalized number of single stars
    with time. Crosses, stars, open, and filled squares denote simulations with initially 30\%, 50\%, 75\%, and 100\% binaries, respectively.}
  \label{fig:evolution_normalized_bf_and_number_binary_systems}
\end{figure}

Observations of binary populations in different environments usually use the binary frequency as the main observable of a binary
population since it can easily be compared between different environments and different observing campaigns. Theoretical studies
of the evolution of binary populations are unaffected by from this shortcoming but usually also use this simple observable for
reasons of comparison with observational data \citep[see for
example][]{1995MNRAS.277.1491K,2001MNRAS.321..699K,2003MNRAS.346..343K,2005MNRAS.358..572I,2007ApJ...665..707H,2008MNRAS.388..307S}. If
we also do this here, we obtain the result described by Fig.~\ref{fig:evolution_normalized_bf_and_number_binary_systems}a. It
shows the evolution of the normalized binary frequency during the first three \Myr\ of cluster evolution in the simulations. The
normalization is performed by dividing the given binary frequencies by the initial binary frequency and allows direct comparison of the
evolution of the binary frequencies for simulations with different initial binary frequencies.

As expected from previous studies \citep{1995MNRAS.277.1491K,2003MNRAS.346..343K,2009MNRAS.397.1577P}, dynamical evolution reduces
the binary frequency with time in all simulations. This can be explained by the destruction of wide binary systems by means of
three body encounters \citep[see][for more details]{1995MNRAS.277.1491K,1995MNRAS.277.1507K}. Apart from this general trend,
Fig.~\ref{fig:evolution_normalized_bf_and_number_binary_systems}a shows that the evolution of the binary population depends on the
initial binary frequency in the sense that the higher the initial binary frequency, the higher the fraction of binaries that are
destroyed during cluster evolution. This finding could lead to the conclusion that the evolution of binary populations is stronger
in clusters with a higher initial binary frequency. An analogous dependence was found by \citet{2008MNRAS.388..307S}, who
investigated the evolution of the binary population in globular star clusters using an analytical approach.

In short, the evolution of the binary population in star clusters seems to depend on the initial binary frequency. However,
because the calculation of the \emph{binary} frequency depends on the number of \emph{single} stars (see
Eq.~\ref{eq:binary_frequency}) it is probably not an unbiased measure of the evolution of the binary population. We thus introduce
the normalized number of binary systems \Nbinnorm\ (see Eq.~\ref{eq:nbin_norm}) as the proper quantity and show its evolution in 
Fig.~\ref{fig:evolution_normalized_bf_and_number_binary_systems}b.

As expected from Fig.~\ref{fig:evolution_normalized_bf_and_number_binary_systems}a, the normalized number of binaries also
decreases with time. However, in contrast to the normalized binary frequency the evolution of the normalized number of
binary systems with time does \emph{not} depend on the initial binary frequency. The difference among the data of only $1\%$ is of
the order of the statistical error in the simulation. How can this difference in the evolution of the binary frequency and the
binary population be explained? It is the evolution of the single star population in the star clusters that significantly affects
the evolution of the binary frequency. Its evolution is show in Fig.~\ref{fig:evolution_normalized_bf_and_number_binary_systems}c.

In contrast to the binary frequency and the number of binaries, the number of single stars in the star cluster obviously increases with time
because the dissolution of a binary system creates two new single stars. The evolution of the normalized number of single stars,
however, depends even more strongly on the initial binary frequency than the evolution of the binary frequency.

We show by means of simple arithmetic why the evolution of the single star population depends on the initial binary frequency
while the evolution of the binary population does not. First, we assume that the evolution of the number of binary systems can be
described by a function $\alpha (t,\binaryfrequency,\dots)$ that may depend on the time of evolution, initial binary frequency,
and other parameters, so that $\Nbin (t) = \alpha (t,\binaryfrequency,\dots) \Nbin(0)$, where $\Nbin (t)$ is the number of binary
systems at time $t$. Dividing this by the initial number of binary systems we obtain the evolution of the normalized number of
binary systems
\begin{equation}
\label{eq:evolution_normalized_number_of_binary_systems}
\Nbinnorm (t) = \frac{\alpha (t,\binaryfrequency,\dots) \Nbin(0)}{\Nbin(0)} = \alpha (t,\binaryfrequency,\dots).
\end{equation}
If we also assume, based on the findings of Fig.~\ref{fig:evolution_normalized_bf_and_number_binary_systems}b, that the evolution
of the normalized number of binary systems does not depend on the initial binary frequency, the function $\alpha$ can be expressed
as $\alpha (t,\binaryfrequency,\dots ) \approx \alpha (t)$. Because the dissolution of a binary system leads to the generation of
two new single stars, the evolution of the number of single stars can be written as
\begin{equation}
\label{eq:evolution_number_of_single_stars}
\Nsing (t) = \Nsing (0) +  2\left[ 1-\alpha (t,\binaryfrequency,\dots) \right] \Nbin (0),
\end{equation}
where the first summand represents the initial number of single stars and the second summand gives the number of newly formed
single stars. Applying the relation $\Nbin (0) = \frac{\binaryfrequency (0)}{1-\binaryfrequency (0)} \Nsing(0)$ and dividing by
the initial number of single stars $\Nsing (0)$, we find that the evolution of the normalized number of binary systems is given by
\begin{equation}
\label{eq:evolution_normalized_number_of_single_stars}
\Nsingnorm (t) = 1 + 2( 1 - \alpha (t) ) \frac{\binaryfrequency (0)}{1 - \binaryfrequency (0)}.
\end{equation}
Hence, in contrast to the normalized number of \emph{binary} systems (Eq.~\ref{eq:evolution_normalized_number_of_binary_systems}),
which depends only on $\alpha$, the normalized number of \emph{single} stars depends explicitly on the initial binary frequency in
the cluster. Consequently, using Eqs.~\ref{eq:evolution_normalized_number_of_binary_systems} and
\ref{eq:evolution_normalized_number_of_single_stars}, we derive an analogous dependence for the normalized binary frequency
\begin{align}
\label{eq:evolution_normalized_number_of_binary_frequency}
^{\binaryfrequency (t)} / _{\binaryfrequency (0)} &= \frac{\Nbin (t)}{\Nsing (t) + \Nbin (t)} \frac{1}{\binaryfrequency (0)} \notag\\
                    &= \frac{\alpha (t)}{\frac{1-\binaryfrequency (0)}{\binaryfrequency (0)} - \alpha (t) + 2} \frac{1}{\binaryfrequency (0)}
\end{align}
In summary, our simple model
(Eqs.~\ref{eq:evolution_normalized_number_of_binary_systems}-\ref{eq:evolution_normalized_number_of_binary_frequency}) describes
the evolution presented in Fig.~\ref{fig:evolution_normalized_bf_and_number_binary_systems} very well: it is the evolution of the
single star population that is very sensitive to the initial binary frequency, while the binary population evolves independently of
the initial binary frequency. The normalized number of binary systems (Eq.~\ref{eq:nbin_norm}) introduced here is therefore the
most simple measure of the evolution of a binary population \emph{independent} of the initial binary frequency.

As a consequence, we can use Eq.~\ref{eq:evolution_normalized_number_of_binary_frequency} to calculate the initial binary
frequency in young ONC-like star clusters from only a few observables. Solving
Eq.~\ref{eq:evolution_normalized_number_of_binary_frequency} for $\binaryfrequency (0)$ reveals that
\begin{equation}
  \label{eq:initial_binary_frequency}
  \binaryfrequency (0) = \frac{\binaryfrequency (t) - \alpha(t) + 1}{\alpha (t)}.
\end{equation}
This finding has a fundamental consequence: if we are able to determine the explicit form of the function $\alpha(t,
\binaryfrequency, \dots)$, we are able to predict the past and future number of binary systems from the present number of binaries
in any cluster environment. If we again assume that $\alpha (t,\binaryfrequency, \dots) = \alpha (t)$, we find that
$\alpha_{\mathrm{fit}}(t) = at^{b} + ct + 1$ with $a=-0.35 \pm 0.04, b=0.89 \pm 0.01 $, and $c =0.27 \pm 0.04 $, provides an
excellent fit to the data in Fig.~\ref{fig:evolution_normalized_bf_and_number_binary_systems}b. However, we emphasize that our
assumptions neglects the potential effects of other parameters such as the stellar density or distribution of binding energy,
which will be investigated in a forthcoming paper.

Nonetheless, for illustrative purposes we assume that the ONC has cluster age of $1\Myr$
\citet{1997AJ....113.1733H} and obtain $\alpha (1\Myr) = 0.92$ using the given fit. Assuming a present-day binary frequency of
$60\%$, motivated by observations that find a consistency between the binary populations of field G dwarfs
\citep{1991A&A...248..485D} and the ONC \citep{1998ApJ...500..825P,2006A&A...458..461K}, we derive an initial binary frequency
of 74\%. This is clearly below the binary frequency observed in young sparse star-forming regions as Taurus-Auriga
\citep{1993A&A...278..129L,1993AJ....106.2005G} and suggests that the initial binary frequency depends on the star-forming
environment as stated by \citet{1994A&A...286...84D}.

We note that this result is based on our specific cluster model, which does not include all the physical processes present in real
star clusters. Thus, further investigations may alter the determined initial binary frequency to some extent. For example,
including the gas the young clusters are embedded in would probably require a higher initial density since star clusters expand
significantly after gas expulsion \citep[e.g.][]{2007MNRAS.380.1589B}. Higher densities should increase the fraction of
dynamically destroyed binary systems during cluster evolution, resulting in larger values of $\alpha$. This means that the ONC
could have been born with a higher initial binary frequency than estimated from our current simulations \citep[as demonstrated
by][]{2001MNRAS.321..699K}.

\begin{figure}
  \centering
  \includegraphics[width=\linewidth]{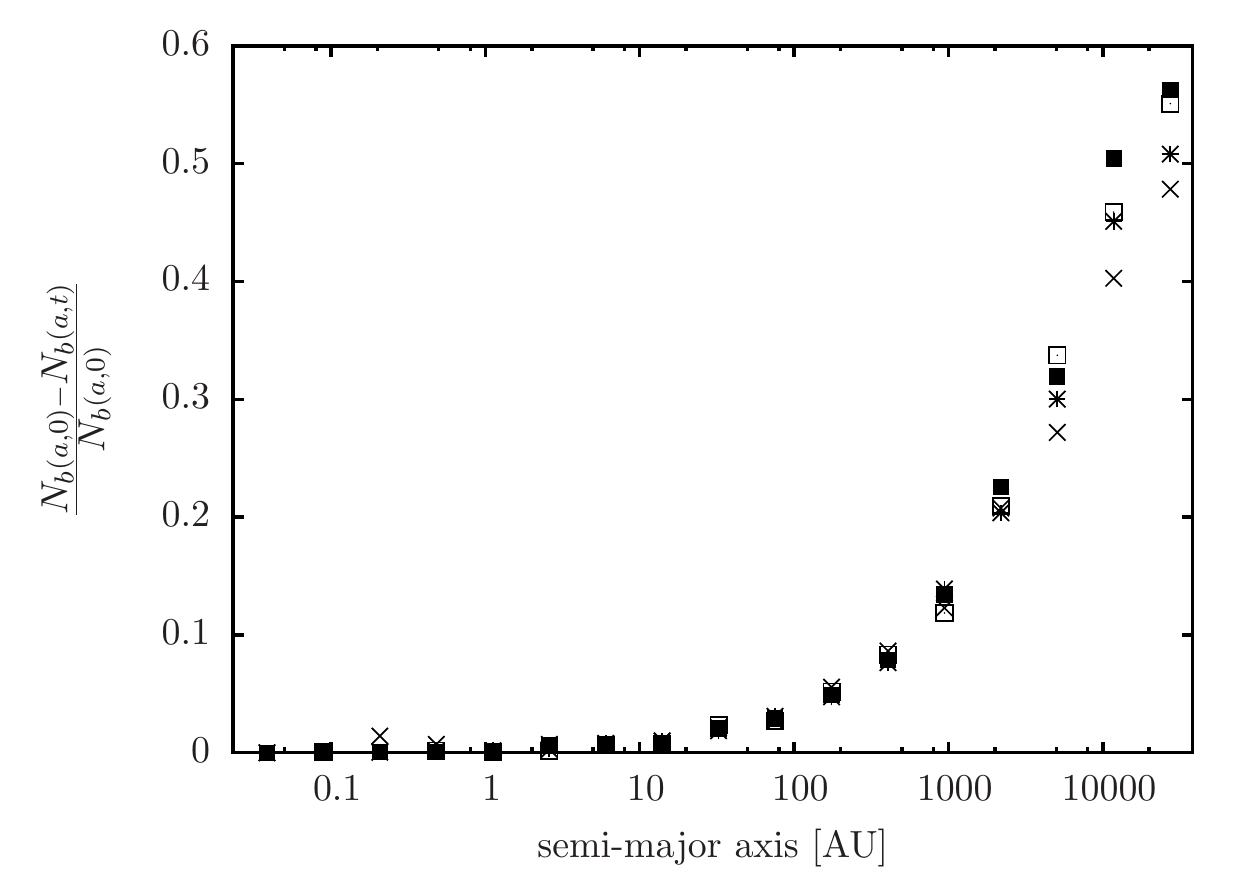}
  \caption{Fraction of binary systems per semi-major axis bin that is destroyed after 1\Myr\ of cluster evolution. Crosses, stars,
    open, and filled squares denote simulations with initially 30\%, 50\%, 75\%, and 100\% binaries, respectively.}
  \label{fig:fraction_number_of_bins_vs_semi}
\end{figure}

We next use our homogeneous data set to investigate why the binary population evolves independently of the initial binary
frequency. Figure~\ref{fig:fraction_number_of_bins_vs_semi} shows the fraction of binary systems per semi-major axis bin that is
destroyed after 1\Myr\ of cluster evolution. It is evident that only binaries with semi-major axis wider than about $\approx
30\AU$ are destroyed during cluster evolution. This dependence on the semi-major axis was previously reported in other studies
of the evolution of binary populations \citep[e.g.][]{1995MNRAS.277.1507K} and can be explained by the outcome of three body
encounters \citep[for details see][]{1975MNRAS.173..729H}: to dissolve a binary system in a three body interaction, the kinetic
energy of the intruding star must exceed the binding energy of the binary system, or, in other words, the velocity of the intruder
must be larger than the orbital velocity of the binary system. With this, we estimate a lower limit to the semi-major axis of
destroyed binaries in the cluster. Following \citet{2003MNRAS.346..343K}, we estimate the ``thermal period'' $\log_{10}
P_{\mathrm{th}} \approx 4.75$ at which destruction still is possible from
\begin{equation}
\label{eq:thermal_period}
\log_{10}P_{\mathrm{th}} = 6.986 + \log_{10} m_{\mathrm{sys}} - 3\log_{10} \sigma_{3\mathrm{D}},
\end{equation}
using the least massive equal-mass system $m_{\mathrm{sys}} = 0.16\Msun$ and assuming a typical relative velocity equal to the
cluster velocity dispersion in our simulations $\sigma_{3D} \approx 3\kms$. Our result, $\log_{10} P_{\mathrm{th}} \approx
4.75\days$, corresponds to a minimum semi-major axis of about $53\AU$.

\begin{figure}
  \centering
  \includegraphics[width=\linewidth]{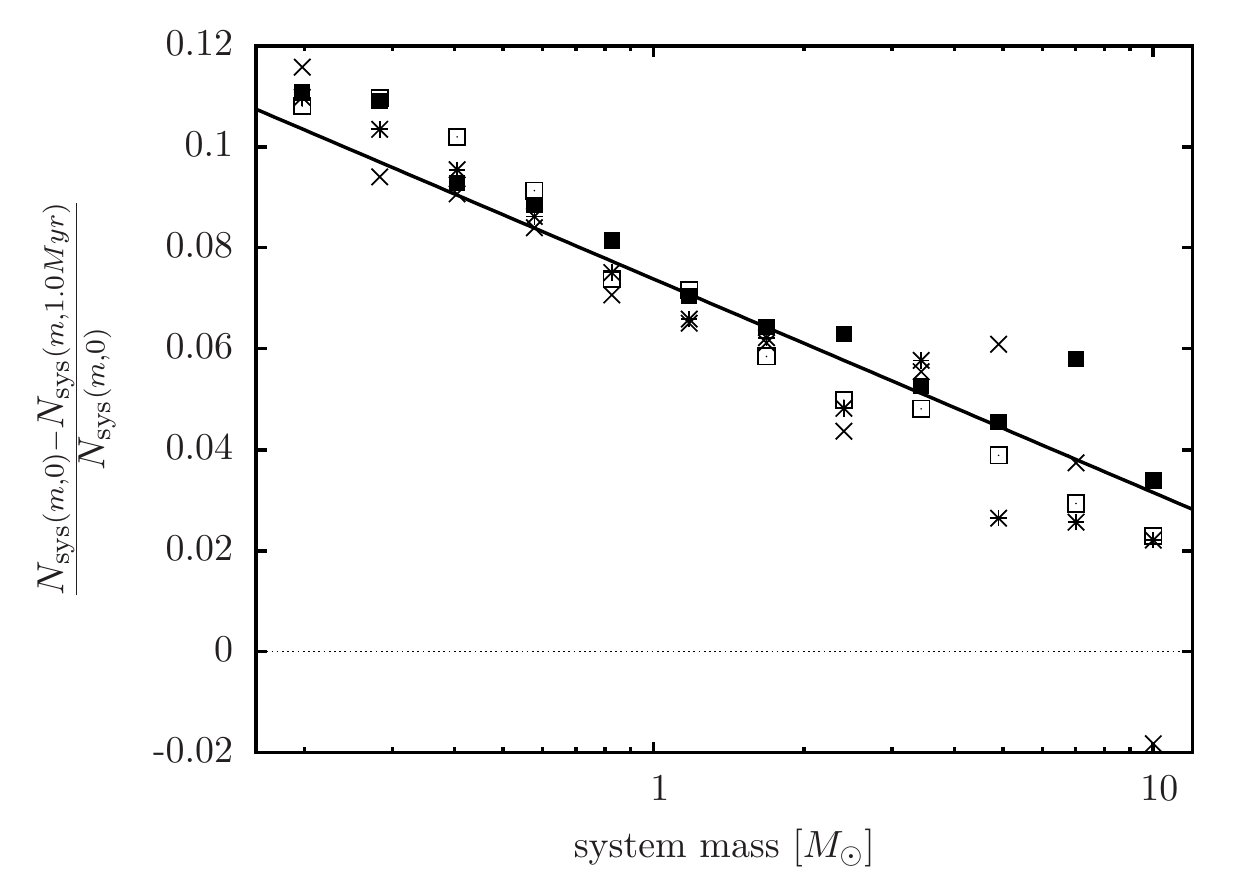}
  \caption{Fraction of binaries per binary system mass bin that are destroyed after 1\Myr\ of cluster evolution. The solid line is
    the fit of the function $f( m_{\mathrm{sys}}) = a \log_{10} m_{\mathrm{sys}} + b$ to the data with initially $75\%$
    binaries. Crosses, stars, open, and filled squares denote simulations with initially 30\%, 50\%, 75\%, and 100\% binaries,
    respectively.}
  \label{fig:fraction_number_bin_sys_vs_sys_mass}
\end{figure}

This estimate is in good agreement with the results of our simulations in Fig.~\ref{fig:fraction_number_of_bins_vs_semi}. However,
in our simulations, there is no sharp transition between stable and destroyed systems because the thermal period $P_{th}$ depends
strongly on the system mass (Eq.~\ref{eq:thermal_period}). Since the periods of the binaries in the simulations have been sampled
independently of the binary system masses the fraction of binaries destroyed by the dynamical evolution of the cluster decreases
with binary mass. This is demonstrated in Fig.~\ref{fig:fraction_number_bin_sys_vs_sys_mass}. The fraction of destroyed binaries
clearly decreases with binary mass.

The solid line in Fig.~\ref{fig:fraction_number_bin_sys_vs_sys_mass} represents a fit of the function $f( m_{\mathrm{sys}}) = a
\log_{10} m_{\mathrm{sys}} + b$ to the data of the simulations with initially $75\%$ binaries in the range from $m_{\mathrm{sys}}
= 0.16\Msun$ up to $10\Msun$. For higher masses, the results suffer from low number statistics and have therefore been excluded
from the fit. We note that $f(m_{\mathrm{sys}})$ also fits the data of the other simulations demonstrating that the evolution of the binary
population is independent of the initial binary frequency.

\begin{figure}
  \centering
\includegraphics[width=\linewidth]{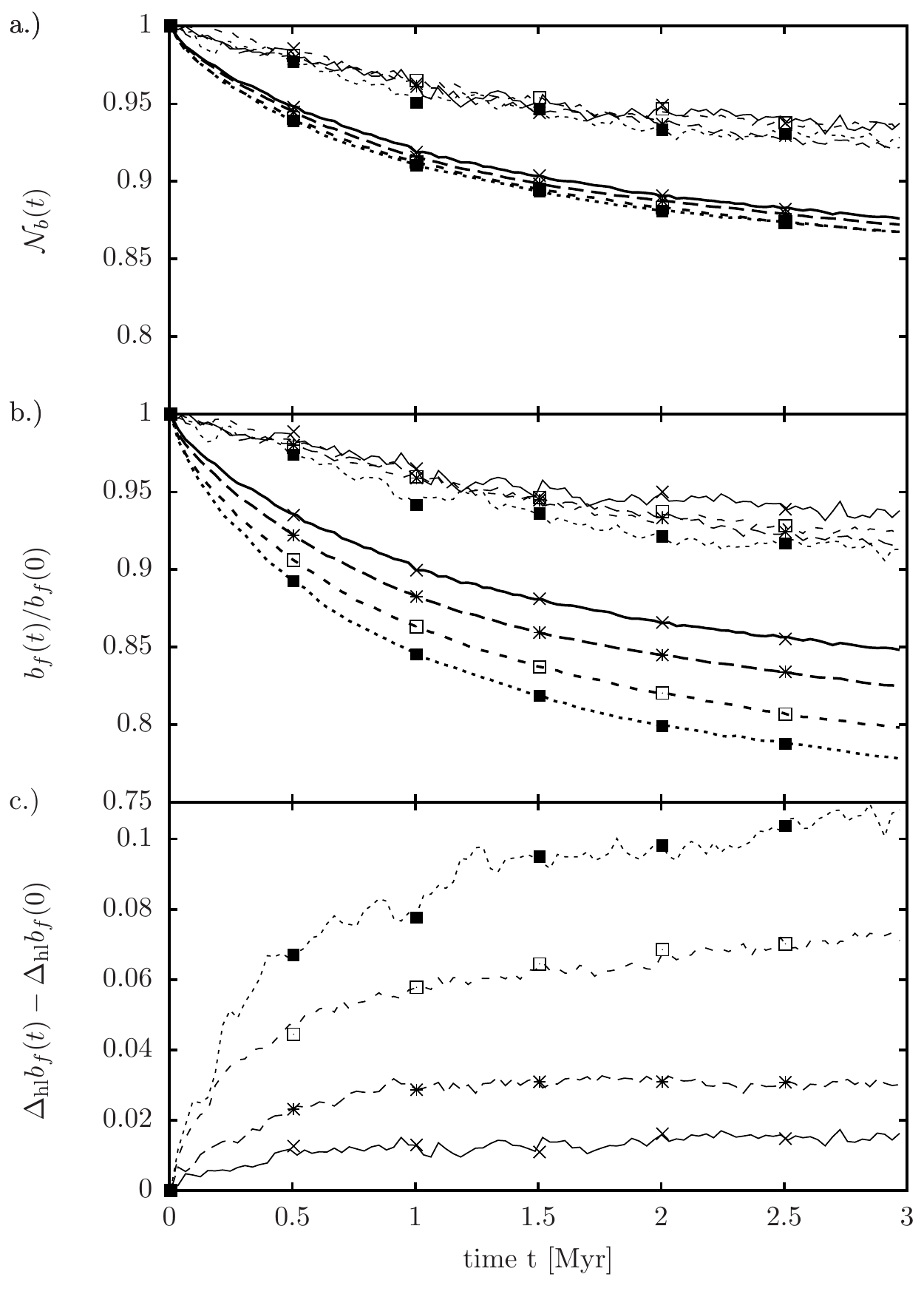}
\caption{Panel a: Evolution of the normalized number of binary systems with primary masses exceeding $2\Msun$ (thin lines)
  and binaries with masses below $2\Msun$ (thick lines) with time. Panel b: Evolution of the normalized binary frequencies of
  binaries with high- (thin lines) and low-mass (thick lines) primaries with time. Panel c: Evolution of the difference of the
  binary frequency of high- and low-mass stars with time. The difference at each time-step has been reduced by the initial
  difference of the binary frequencies to improve the readability. Crosses, stars, open, and filled squares denote simulations with
  initially 30\%, 50\%, 75\%, and 100\% binaries, respectively.}
  \label{fig:normalized_bin_freq_vs_time__masses}
\end{figure}

The strong mass dependence of the binary destruction is further illustrated in
Fig.~\ref{fig:normalized_bin_freq_vs_time__masses}a. It shows the evolution of the normalized number of binaries with primary
masses either exceeding $2\Msun$ (thin lines) and lower than $2\Msun$ (thick lines) in clusters with various initial binary
frequencies. Clearly, a larger fraction of low-mass binaries is destroyed during the dynamical evolution of the cluster.

The boundary of $2\Msun$ between high and low-mass primaries was chosen in accordance with the study of \citet{2006A&A...458..461K},
who performed a high spatial resolution infrared study of the binary population in the ONC. They found the frequency of binaries
with high-mass primaries to be significantly higher. Figure~\ref{fig:normalized_bin_freq_vs_time__masses}a implies that this finding
can be explained by the dynamical evolution of the cluster and does not depend on the initial binary frequency.

However, a consistent comparison of our simulation and the observations by \citet{2006A&A...458..461K} must focus on the frequency
and not the number of binaries. Figure~\ref{fig:normalized_bin_freq_vs_time__masses}b shows the evolution of the frequency of
binaries with high and low-mass primaries using the same symbols as in
Fig.~\ref{fig:normalized_bin_freq_vs_time__masses}a. Clearly the binary frequencies of the high and low-mass primaries evolve
quite differently. The former decreases much more slowly with time than the latter and
does not depend on the initial binary frequency. In contrast, the evolution of binaries with low-mass primaries \emph{does} depend
on the initial binary frequency.

The differences in the evolution are explained by two mechanisms. First, binaries with low-mass primaries are preferentially
destroyed by dynamical interactions in the cluster because of their lower binding energy
(Fig.~\ref{fig:normalized_bin_freq_vs_time__masses}a). Secondly, the destruction of a binary by a high-mass primary can result
in the generation of two new single high-mass stars or a high and a low-mass star. The latter possibility means that the number of
single low-mass stars increases, hence the frequency of binaries with low-mass primaries decreases relative to the
former case. Simultaneously, the frequency of binaries with a high-mass primary does not decrease as fast as implied by
Eq.~\ref{eq:evolution_normalized_number_of_binary_frequency} because of the smaller increase in the number of single high-mass
stars. Hence, the possible generation of a high- and a low-mass star in the destruction of a binary with a high-mass primary
accelerates the reduction of the binary frequency of low-mass stars and decelerates the reduction of the binary frequency of
high-mass stars.

We emphasize the different evolution of binaries with high and low-mass primaries in
Fig.~\ref{fig:normalized_bin_freq_vs_time__masses}c. Here we show the difference between the binary frequencies of high and low
mass stars over time, $\Delta_{\mathrm{hl}} \binaryfrequency (t)$, relative to its initial value $\Delta_{\mathrm{hl}} (t)$. This
normalization accounts for the possible initial difference of the binary frequencies of high and low-mass stars, so that all values
are zero initially.

In all simulations, the difference of the binary frequency of high and low-mass stars increases with time because the binary
frequency of low-mass stars decreases faster than the binary frequency of high-mass stars
(Fig.~\ref{fig:normalized_bin_freq_vs_time__masses}b). However, the amount of increase depends strongly on the initial binary
frequency in the cluster with the lowest initial binary frequency ($30\%$) resulting in the smallest increase of less than $2\%$ after
3\Myr\, while in simulations with the maximum initial binary frequency ($100\%$) the difference increases by more than
$10\%$. Under the assumption that initially the frequency of binaries with low- and high-mass stars was the same in the ONC, the
observational data from \citet{2006A&A...458..461K} imply that the binary frequency in the ONC initially was very high with values
well above $75\%$. If the initial binary frequencies were lower initially the observed difference must have been part of the
initial conditions of the binary population.

% %
% %______________________________________________________________

\section{Conclusions}
\label{sec:conclusions}

We have studied the dynamical evolution of the binary population in young dense ONC-like star-clusters using two different
measures: the (normalized) binary frequency (Eq.~\ref{eq:binary_frequency}) and the normalized number of binary systems
(Eq.~\ref{eq:nbin_norm}). As expected, the evolution of the binary frequency depends heavily on the initial binary frequency: the
higher the initial binary frequency in a cluster, the stronger the decrease in the binary frequency during cluster evolution. However,
our main result is that the evolution of the normalized number of binaries is independent of the initial binary frequency. During
the first \Myr\ of cluster evolution, the number of binary systems decreases by about $6-9\%$ independent of the initial binary
frequency in the clusters.

The reason for this discrepancy between the two measures is the dependence of the binary frequency on the population of single
stars. Since the destruction of a binary system leads to the generation of two new single stars, the single star population can
easily dominate the evolution of the binary frequency in a cluster, especially in the case of high initial binary frequencies
(Eq.~\ref{eq:evolution_normalized_number_of_binary_frequency}). Hence, the number of binaries -- unbiased by the evolution of the
single star population -- represents the proper measure of the evolution of the binary population. We can therefore conclude that
the evolution of the \emph{binary population} in ONC-like stars clusters is \emph{independent} of the initial binary frequency,
which is in striking contrast to what would be inferred from the evolution of the binary frequency.

This independence of the number of binaries has allowed us to determine the initial binary frequency and predict the evolution of
the binary population in young ONC-like stars clusters from present-day observational data. For example, taking the observed
binary frequency of the ONC of $\sim 60\%$ and its age of $1\Myr$ \citep{1998ApJ...500..825P,1997AJ....113.1733H}, and estimating
$\alpha (1\Myr) = 0.92$ from our fit to the simulations, Eq.~\ref{eq:initial_binary_frequency} provides an initial binary
frequency of $74\%$. This binary frequency is significantly lower than the binary frequency observed in sparse star-forming
regions such as Taurus-Auriga \citep[e.g.][]{1993A&A...278..129L,1993AJ....106.2005G} and implies that the initial binary
frequency depends on the star-forming region. This finding agrees well with the results of \citet{1994A&A...286...84D}, who
claimed that the primordial binary frequency in star-forming regions depends on the temperature of the natal molecular cloud in
which the stars are forming of.  However, different initial conditions of the clusters (e.g. higher initial stellar densities)
could possibly increase the amount of destroyed binary systems, which would further increase the estimated initial binary frequency
\citep{2001MNRAS.321..699K}. It is not trivial to quantify these effects, so further numerical studies are required.

In addition, our simulations show that the evolution of the binary populations depends crucially on the masses of the primary
stars: the lower the mass of the primary, the more likely that the binaries are destroyed. Over time, this leads to a higher
binary frequency for stars with masses exceeding $2\Msun$ than for lower mass stars. This is even so in the case where the binary
frequencies of high and low-mass stars were initially the same. This finding is in good agreement with the observations of
\citet{2006A&A...458..461K}, who found a significantly higher binary frequency for high-mass stars than for low-mass stars in the
ONC. Our results suggest that the difference in the binary frequencies of high- and low-mass stars are probably not the outcome of
the binary formation process but rather the result of the dynamical evolution with time. A possible test of this hypothesis would
be to compare the maxima of the period/semi-major axis distributions of high- and low-mass stars in young dense star
clusters. Since the probability of destroying a binary in a three body encounter depends on the binding energy and therewith on the
separation and mass of the binary, the maximum of the period and semi-major-axis distributions for massive stars should be at longer
periods and wider semi-major axis than for lower mass stars. However, this test is not easy to be carried out since observations of
the binary population in young dense star clusters usually only cover a small range of periods/separations of the binaries, and
it is often not even possible to determine the physical orbital parameters based on the observations.

The simulations presented here were performed for ONC-like clusters only and used the specific binary population model of
\citet{2007A&A...474...77K}. We aim to generalise the present results by performing simulations with different cluster and
binary population models for an forthcoming paper.

%
%______________________________________________________________

\begin{acknowledgements}
Simulations were performed at the J\"ulich Supercomputer Centre, Research Centre J\"ulich, Project HKU14.
\end{acknowledgements}

% Bibliography.
% -------------
\bibliographystyle{aa}
\bibliography{reference}

% \begin{thebibliography}{}

% \end{thebibliography}

\end{document}